\documentclass[twocolumn,prd,superscriptaddress,floatfix]{revtex4}

\usepackage{apjfonts}
\usepackage{bm}
\usepackage{graphicx}
\usepackage{subfigure}

\def\be{\begin{equation}}
\def\ee{\end{equation}}
\def\bea{\begin{eqnarray}}
\def\eea{\end{eqnarray}}
\def\nn{\nonumber}
\def\gm{{\rm (geom)}}
\def\gw{{\rm (grow)}}

\begin{document}

\title{Is Modified Gravity Required by Observations? \\
An Empirical Consistency Test of Dark Energy Models}

\author{Sheng Wang}
\affiliation{Brookhaven National Laboratory, Upton, NY 11973--5000, USA}
\affiliation{Department of Physics, Columbia University, New York, NY 10027, USA}
\author{Lam Hui}
\affiliation{Department of Physics, Columbia University, New York, NY 10027, USA}
\affiliation{Institute for Strings, Cosmology and Astroparticle Physics, Columbia University, New York, NY 10027, USA}
\author{Morgan May}
\affiliation{Brookhaven National Laboratory, Upton, NY 11973--5000, USA}
\author{Zolt\'{a}n Haiman}
\affiliation{Department of Astronomy, Columbia University, New York, NY 10027, USA}

\date{\today}

\begin{abstract}
We apply the technique of parameter splitting to existing cosmological
data sets, to check for a generic failure of dark energy models.  Given
a dark energy parameter, such as the energy density $\Omega_\Lambda$ or
equation of state $w$, we split it into two meta--parameters with one
controlling geometrical distances, and the other controlling the growth
of structure.  Observational data spanning Type Ia supernovae, the
cosmic microwave background (CMB), galaxy clustering, and weak
gravitational lensing statistics are fit without requiring the two
meta--parameters to be equal.  This technique checks for inconsistency
between different data sets, as well as for internal inconsistency
within any one data set (\textit{e.g.}, CMB or lensing statistics) that
is sensitive to both geometry and growth.  We find that the cosmological
constant model is consistent with current data.  Theories of modified
gravity generally predict a relation between growth and geometry that is
different from that of general relativity.  Parameter splitting can be
viewed as a crude way to parametrize the space of such theories.  Our
analysis of current data already appears to put sharp limits on these
theories: assuming a flat universe, current data constrain the
difference $\Delta\Omega_\Lambda = \Omega_\Lambda\gm-\Omega_\Lambda\gw$
to be $-0.0044^{+0.0058+0.0108}_{-0.0057-0.0119}$ ($68\%$ and $95\%$
C.L. respectively);  allowing the equation of state $w$ to vary, the
difference $\Delta w = w\gm - w\gw$ is constrained to be
$0.37^{+0.37+1.09}_{-0.36-0.53}$.  Interestingly, the region
$w\gw > w\gm$, which should be generically favored by theories that slow
structure formation relative to general relativity, is quite restricted
by data already.  We find $w\gw < -0.80$ at $2\sigma$.  As an example,
the best--fit flat Dvali--Gabadadze--Porrati (DGP) model approximated by
our parametrization lies beyond the $3\sigma$ contour for constraints
from all the data sets.
\end{abstract}

\maketitle

%%%
\section{Introduction}

Observations of distant supernovae (SNe), galaxies, clusters of
galaxies, and the cosmic microwave background (CMB) have shown that,
surprisingly, the cosmic expansion is accelerating.  This reveals that
fundamentally new physics is missing from our understanding of the
universe~\cite{DETF}.

The cosmic acceleration may arise either from ``dark energy,'' a
mysterious yet presently dominant component of the total energy density,
or from ``modified gravity,'' a modification of general relativity (GR)
on large scales.  The first case includes, for example, Einstein's
cosmological constant or quintessence, a dynamical scalar
field~\cite{Quint}.  The second case includes modifications of
four--dimensional GR due to the presence of extra dimensions, 
scalar--tensor theories, and others~\cite{DGP,MG,others}.

Current efforts focus, within the dark energy paradigm, on improving the
constraints on the dark energy density $\Omega_{\rm DE}$, its equation
of state (EOS) $w\equiv P/\rho$ and its time evolution $dw/da$ (where
$a$ is the scale factor), by using observational data that bear on
geometrical distances and the growth of structure.  As first emphasized
by~\cite{LSS} and subsequently discussed by many others~\cite{many}, GR
predicts a definite relation between geometrical distances and growth
which is generically violated by modified theories of gravity.  To the
extent current data (that are sensitive to different combinations of
geometry and growth) yield consistent dark energy constraints, one can
interpret this as a confirmation of the dark energy + GR framework.  The
simplest dark energy model, the cosmological constant, has passed this
kind of consistency test so far~\cite{WMAP}.

In this paper, we sharpen the consistency test.  Our method goes by the
name of  ``parameter splitting'' as proposed by~\cite{ZHS,CK}.  Let us
illustrate the technique using the cosmological constant
($\Lambda$)--cold dark matter (CDM) model.  Instead of
fitting the suite of observational data with a single cosmological
constant density parameter $\Omega_\Lambda$ (in addition to, of course,
other non--dark energy parameters), we fit them with two parameters
$\Omega_\Lambda\gm$ and $\Omega_\Lambda\gw$: one determining the
geometrical distances, and the other controlling the growth of
structure.  The conventional approach is to assume the two parameters
are equal.  Here, they are allowed to vary separately.  We employ the
Markov chain Monte Carlo (MCMC) technique~\cite{MCMC} to derive the
marginalized constraints on both parameters.  If the $\Lambda$CDM model
is correct, these two parameters should agree within their
uncertainties.  This technique of splitting a conventional parameter
into two ``meta--parameters'' can of course be applied to any other
parameter.  In this paper, we will consider the splitting of both
$\Omega_\Lambda$ and $w$.

It is important to emphasize that parameter splitting checks for
consistency not only between different data sets, but also for internal
consistency within any single data set that is sensitive to both
geometry and growth.  In some sense, the conventional approach of
obtaining constraints on, \textit{e.g.}, $\Omega_\Lambda$ separately from
SNe, CMB, lensing and so on, and checking that they are consistent, is
itself a simple form of parameter splitting, \textit{i.e.}, splitting
$\Omega_\Lambda$ into $\Omega_\Lambda$(SNe), $\Omega_\Lambda$(CMB),
$\Omega_\Lambda$(lensing), etc.  The parameter splitting that we employ
here represents a more stringent, and theoretically better motivated,
consistency test.  It is also useful to note that there is a wide
variety of modified gravity theories.  Our splitting of $\Omega_\Lambda$
and $w$ can be thought of as a crude way to parametrize the space of
such theories.  For instance, in the Dvali--Gabadadze--Porrati (DGP)
theory~\cite{DGP} where gravity becomes weaker on large scales,
structure growth is slowed and therefore one expects qualitatively
$w\gw > w\gm$~\cite{LSS,footnote1}.

We caution that should an inconsistency be discovered via
parameter splitting, modified gravity is not the only possible
interpretation.  Systematic problems with the data, as well as
complications in the dark energy model (such as a time varying $w$ or
nontrivial dark energy clustering~\cite{kunz}), are also possible.
Additional parameters need to be introduced to check for the latter
case.  Parameter splitting can be applied to the new parameters as
appropriate.

%%%
\section{Geometry}

All geometrical distances in cosmology, such as the luminosity or
angular diameter distance, are related to the radial comoving distance
\be
\chi(z) = \int^z_0 \frac{dz'}{H(z')},
\label{eq:chi}
\ee
setting the speed of light $c=1$.  The Hubble parameter $H$ as a
function of redshift $z$, \textit{i.e.}, the expansion history, can be
parametrized as follows:
\be
\frac{H^2(z)}{H^2_0} = \Omega_m (1+z)^3 + \Omega_r (1+z)^4
+ \Omega_{\rm DE} (1+z)^{3(1+w)},
\label{eq:hist}
\ee
where $H_0=100h$ km s$^{-1}$Mpc$^{-1}$ is the Hubble constant today.
Throughout this paper, we assume that the universe is spatially flat,
the dark energy has a constant EOS parameter $w$ and all three species
of neutrinos are massless.  $\Omega_r$ is the radiation density today,
in units of the critical density, including photons and massless
neutrinos; $\Omega_{\rm DE}$ is the present dark energy density, denoted
as $\Omega_\Lambda$ for the cosmological constant model ($w=-1$).  Note
that for a flat universe, the dimensionless matter density $\Omega_m$
can be replaced by $1-\Omega_r-\Omega_{\rm DE}$.  We will use a
superscript ``(geom)'' to denote the dark energy parameters appearing in
the expressions of geometrical distances.

%%%
\section{Growth}

Inhomogeneities grow under gravitational instability according to the
prevailing structure formation paradigm.  The dynamics within the GR
framework is described by a set of Boltzmann--Einstein equations well
documented in the literature~\cite{dodelson}.  In this paper, we use the
publicly available code \textit{CAMB}~\cite{LCL} to evolve these
equations.  For the purpose of illustrating our method, and purely for
this purpose, let us consider the special case of subhorizon matter
fluctuations in the late universe.  They evolve according to
$\ddot{\delta}_m + 2H\dot{\delta}_m = 4\pi G\rho_m\delta_m$, where
$\delta_m\equiv\delta\rho_m/\rho_m$ is the matter overdensity, $\rho_m$
is the average matter density, $G$ is the Newton constant and the dots
denote proper time derivatives.  We ignore the dark energy perturbations
here for simplicity.  The growth equation can be rewritten as
\be
\frac{d^2\delta_m}{d\ln{a}^2} + \left[\frac{d\ln{H}}{d\ln{a}} + 2\right]\frac{d\delta_m}{d\ln{a}}
= \frac{3 \Omega_m H_0^2}{2 a^3 H^2} \delta_m,
\label{eq:grow}
\ee
where $a=1/(1+z)$ is the scale factor.  Therefore, the expansion history
[Eq.~(\ref{eq:hist})] that determines geometrical distances also
determines the growth of structure, in a way that is uniquely predicted
by GR.

It is not surprising that, in order to match existing data, viable
theories of modified gravity often predict an expansion history (and
therefore geometrical distances) that is similar to the one in
Eq.~(\ref{eq:hist}).  Such theories, however, generally predict a
relation between expansion history and growth that is different from the
one in Eq.~(\ref{eq:grow}).  Given the wide variety of these theories,
and in the absence of a particularly compelling candidate~\cite{ghosts},
a crude way to test for such a possibility is to allow the dark energy
parameters to take different values in the growth equation
[Eq.~(\ref{eq:grow})] from their values in the expression for distance
[Eq.~(\ref{eq:chi})], \textit{i.e.}, parameter splitting.  We
use a superscript ``(grow)'' to denote the dark energy parameters
characterizing the evolution of inhomogeneities.

Note that one has some freedom in exactly how the parameter splitting
is performed.  For instance, in Eq.~(\ref{eq:grow}), the dark energy
parameters show up in two places: the second term on the left hand side
of the equation ($d\ln{H}/d\ln{a}$) and the term on the right hand side
($\Omega_m/H^2$).  One could choose to assign all of them to the
``growth'' category which is what we do, or one could assign some to the
``geometry'' category and the others to the ``growth'' category.
Ultimately, there are many possible consistency tests, and here we have
chosen to perform one that is particularly simple to implement,
\textit{i.e.}, assigning all dark energy parameters that enter the
fluctuation equations to the ``growth'' category.  It is worth
noting that in a lot of modified gravity theories, the equivalent of the
Poisson's equation is often modified without modifying energy--momentum
conservation.  In that case, one could argue assigning the term on the
right hand side of Eq.~(\ref{eq:grow}) alone to the ``growth'' category
might make more sense.  We hope to investigate this in the future.

The exact Boltzmann--Einstein equations for the evolution of structure,
allowing for multiple components, photons, neutrinos and so on, are more
complicated than Eq.~(\ref{eq:grow}).  The same parameter--splitting
scheme can nevertheless be applied to the exact equations, which is what
we do.  This means, for example, the shape of the transfer function,
such as the radiation--matter equality peak of the power spectrum, is
determined by the growth parameters -- recall that the transfer function
is completely determined by the dynamics of fluctuation growth.  The
conversion of a feature, such as the radiation--matter equality length
scale to an observed angle, on the other hand, involves the geometry
parameters.

%%%
\section{The Parameter--Splitting Technique}

To illustrate how the splitting of dark energy parameters into the
``geometry'' and ``growth'' categories is done in our analysis, we start
with the weak lensing (WL) observables.  There exists a natural division
between the two categories for each term involved in the
calculation~\cite{ZHS}.

WL surveys measure the aperture mass statistic on different angular
scales $\theta$:
\be
\langle M^2_{\rm ap}(\theta)\rangle = \frac{1}{2\pi}
\int\ell d\ell~P_\kappa(\ell)W^2(\ell\theta),
\ee
where $W$ is a window function with no dependence on cosmology.
$P_\kappa(\ell)$ is the convergence power spectrum at the angular
wavenumber $\ell$, given by
\bea
\nn
P_\kappa(\ell) &=& \frac{9}{4}\Omega^2_m H_0^4 \int^{\infty}_0 dz
~(1+z)^2 \left[\frac{d\chi(z)}{dz}\right] \xi^2(z)
~P_\delta \left[\frac{\ell}{\chi(z)},z\right], \\
\xi(z) &=& \int^{\infty}_z dz'~n_{\rm gal}(z')
\left[\frac{\chi(z')-\chi(z)}{\chi(z')}\right].
\eea
Here $P_\delta[\ell/\chi,z]$ is the matter power spectrum at wavenumber
$k=\ell/\chi$ and redshift $z$, $n_{\rm gal}$ is the normalized redshift
distribution of the background galaxies, and we have used Limber's
approximation.  We express everything in terms of the redshift $z$,
which is an observable of the surveys.

Consider for instance the splitting of $\Omega_\Lambda$ for the flat
$\Lambda$CDM model.  The three--dimensional matter power spectrum $P_\delta$ and the
mean matter density $\Omega_m$ ($=1-\Omega_\Lambda$, where the
contribution of radiation is neglected at low redshifts) sitting outside
the integral both describe the foreground inhomogeneities through which
photons travel.  Therefore they go into the ``growth'' category and are
calculated using $\Omega_\Lambda^\gw$.  All $\chi$'s within the integral
fall naturally in the ``geometry'' category.  This includes the $\chi$
in the wavenumber $\ell/\chi$, which reflects the conversion between the
observed angle and the physical length scale.  These geometrical
distances are all calculated using $\Omega_\Lambda^\gm$.  A similar
split can be applied to $w$ in the context of the quintessence (Q)--CDM
model.

With the WL example in mind, we next consider the CMB.  The temperature
anisotropy power spectrum is given by
\be
C^{TT}_\ell = \frac{2}{\pi}\int k^2 dk~P_\Psi(k)
\left|\frac{\Theta_\ell(k,z=0)}{\Psi(k)}\right|^2,
\ee
where $\Psi(k)$ is the primordial metric perturbation (in conformal
Newtonian gauge), $P_\Psi(k)\propto k^{n_s-4}$ is the power spectrum of
$\Psi$, and $\Theta_\ell$ is given by~\cite{dodelson}
\be
\Theta_\ell(k,z=0) = \int_0^\infty dz'~\tilde{S}_T(k,z') j_\ell[k \chi(z')].
\label{eq:CMB}
\ee
where $j_\ell$ is the spherical Bessel function and $\tilde{S}_T$
denotes some source function.  All the complicated dynamics is contained
in $\tilde{S}_T$.  Publicly available Boltzmann codes~\cite{SM,LCL} can
be used to compute $\tilde{S}_T$, and therefore $\Theta_\ell$, for any
given primordial perturbation $\Psi$ ($\Theta_\ell/\Psi$ is independent
of $\Psi$; see~\cite{dodelson}).

We perform the geometry--growth split of Eq.~(\ref{eq:CMB}) as
follows~\cite{footnote2}:
$\tilde{S}_T$ falls under the ``growth'' category and the rest (namely
$\chi(z')$ in the argument of $j_\ell$) falls under the ``geometry''
category~\cite{footnote3}.  The rationale
for this particular way of splitting is most transparent when
considering the Sachs--Wolfe term~\cite{SW}, where $\tilde{S}_T(k,z)$ is
well approximated by $\delta_D(z-z_\ast)[\Theta_0+\Psi](k,z_\ast)$.
Here $\delta_D(z-z_\ast)$ is the Dirac delta function with $z_\ast$
being the redshift of last scattering, and $\Theta_0(k,z_\ast)$ and
$\Psi(k,z_\ast)$ are the temperature monopole and metric perturbations
at last scattering.  Therefore, the Sachs--Wolfe term is
\be
\Theta^{\rm SW}_\ell(k,z=0) \simeq \left[\Theta_0+\Psi\right](k,z_\ast)~j_\ell(k\chi_\ast),
\ee
and our geometry--growth split is equivalent to using the growth
parameters to compute $[\Theta_0+\Psi](k,z_\ast)$ and the geometry
parameters to compute $\chi_\ast$, the distance to last scattering.

It is straightforward to generalize the above splitting scheme to
similar expressions describing the polarization spectrum.  In the case
of SNe, parameter splitting is trivial since SNe constrain only the
geometry parameters.  The splitting for galaxy clustering is done as
follows.  As discussed earlier, the growth (as opposed to geometry)
parameters determine the transfer function for the matter power
spectrum.  On the other hand, to measure the three--dimensional power spectrum of
galaxies as a function of comoving spatial scale, one has to adopt a
cosmological model in order to convert the observed redshifts and
angular separations into comoving distances.  This conversion is trivial
for low--redshift surveys (involving only $H_0$) such as the Two--Degree
Field Galaxy Redshift Survey (2dFGRS), but is nontrivial for moderate
redshift samples, such as the luminous red galaxies (LRGs) in the Sloan
Digital Sky Survey (SDSS).  For the LRGs, we follow~\cite{LRG} and
include a cosmology--dependent rescaling of the $k$--axes~\cite{HH}.
This rescaling is included in the ``geometry'' category.

%%%
\section{Current Observations}

Below we list the four data sets used in our analysis.  Many of these,
though not all, are included in the \textit{CosmoMC} package~\cite{CosmoMC}.

\subsection{Cosmic Microwave Background}

We use (i) the recent Wilkinson Microwave Anisotropy Probe (WMAP)
three--year data set~\cite{WMAP}, and (ii) small scale CMB observational
data including Arcminute Cosmology Bolometer Array Receiver
(ACBAR)~\cite{ACBAR}, Balloon Observations Of Millimetric Extragalactic
Radiation and Geophysics (BOOMERanG)~\cite{Boom} and Cosmic Background
Imager (CBI)~\cite{CBI}.  We modify the Boltzmann code \textit{CAMB}~\cite{LCL}
by splitting the dark energy parameters as described above.  We assume
adiabatic initial fluctuations, and neglect B--mode polarization and
tensor modes.

\subsection{Supernovae}

We use the SNe data set for the Supernova Legacy Survey (SNLS) analysis
described in~\cite{SNLS}.

\begin{figure}%[htp]
  \centering
  \subfigure{\includegraphics[width=85mm]{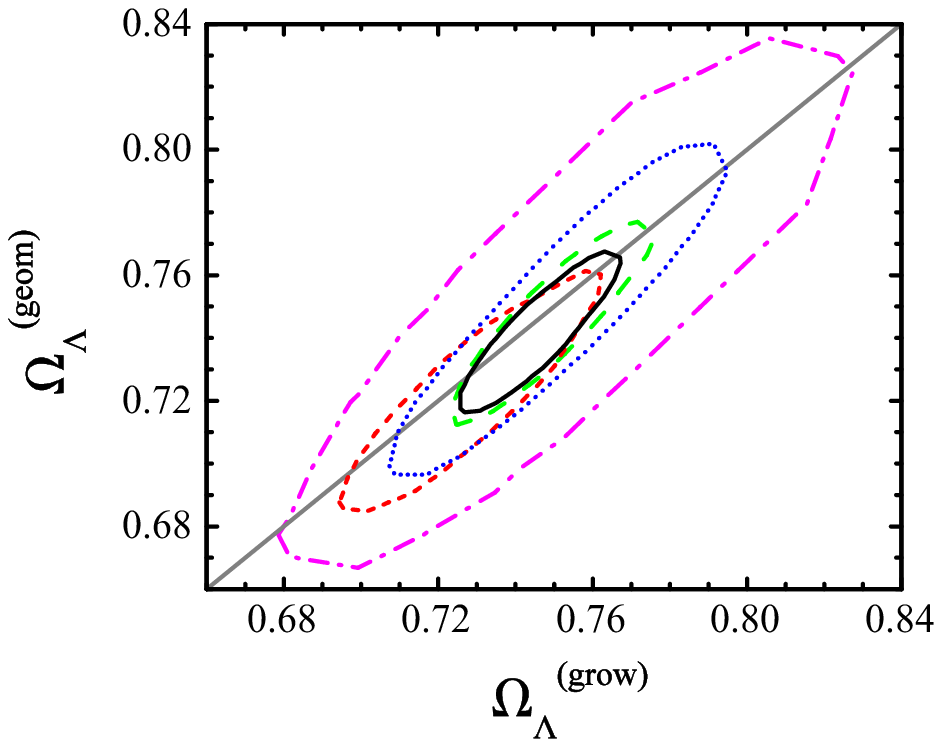}} \\
  \vspace{-1\baselineskip}
  \subfigure{\includegraphics[width=85mm]{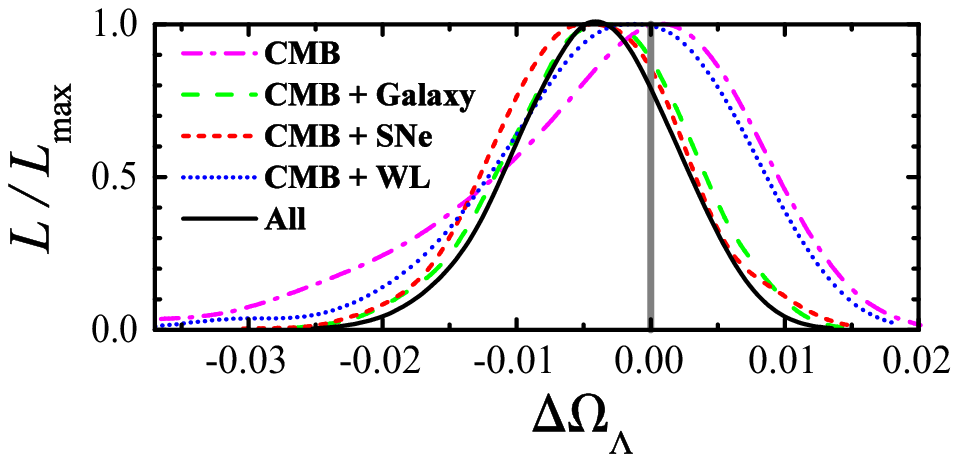}}
  \vspace{-1\baselineskip}
  \caption{\label{fig:LCDM}Joint constraints on $\Omega_{\Lambda}^\gm$
    and $\Omega_{\Lambda}^\gw$ in a $\Lambda$CDM model
    (\textit{upper panel}) and the normalized likelihood distribution of
    $\Delta\Omega_\Lambda\equiv\Omega_\Lambda^\gm-\Omega_\Lambda^\gw$
    (\textit{lower panel}).  Here the equation of state parameters are
    fixed as $w^\gm=w^\gw=-1$.  The contours and curves show the $68\%$
    confidence limits from the marginalized distributions.  The thick
    gray lines show $\Omega_{\Lambda}^\gm=\Omega_{\Lambda}^\gw$.  The
    data sets used are described in the text.  Different contours and
    curves represent constraints from different combinations of the data
    sets.  The smallest contour and the most narrow curve (black solid
    line) represent constraints from all the data.  No significant
    difference is found and deviations are constrained to
    $\Delta\Omega_\Lambda=-0.0044^{+0.0058+0.0108}_{-0.0057-0.0119}$
    ($68\%$ and $95\%$ C.L.).}
\end{figure}
\subsection{Galaxy Clustering}

We use data sets from (i) the Two--Degree Field Galaxy Redshift Survey
(2dFGRS)~\cite{2dF}, which probes the galaxy distribution at redshift
$z\sim 0.1$ and the power spectrum on scales of
$0.022h$ Mpc$^{-1}<k<0.18h$ Mpc$^{-1}$, and (ii) the luminous red
galaxies in the Sloan Digital Sky Survey (SDSS)~\cite{LRG}, which are at
an effective redshift of $z\sim 0.35$ and cover scales between
$0.012h$ Mpc$^{-1}<k<0.20h$ Mpc$^{-1}$.  Redshift--space distortions,
galaxy biasing and nonlinear clustering~\cite{footnote4} are dealt with
in ways described in~\cite{2dF,LRG}.

\begin{figure}%[htp]
  \includegraphics[width = 85mm]{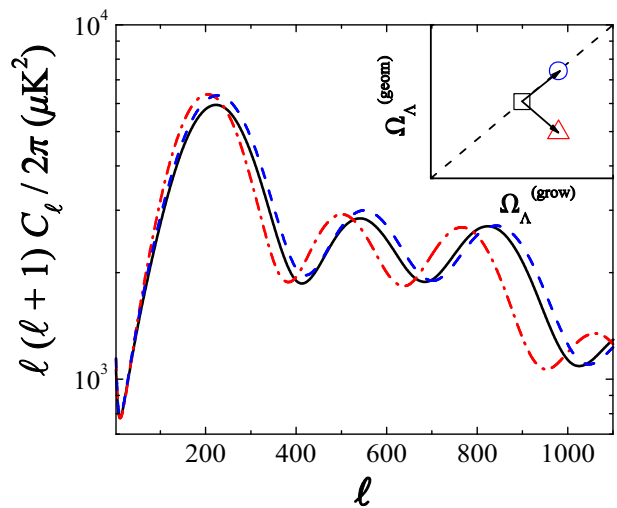}
  \vspace{-1\baselineskip}
  \caption{\label{fig:dif_avg}Variations of CMB temperature power
    spectra due to different changes of $\Omega_\Lambda^\gm$ and
    $\Omega_\Lambda^\gw$ (with all the other cosmological parameters
    fixed) as illustrated in the inset on the
    $\Omega_\Lambda^\gm$ \textit{vs.} $\Omega_\Lambda^\gw$ plane.  The
    black solid curve corresponds to the black square symbol,
    which is our best--fit $\Lambda$CDM model with
    $\Omega_\Lambda^\gm=\Omega_\Lambda^\gw=0.744$.  The blue dashed curve
    corresponds to the blue circular symbol, which is obtained from the
    best--fit model by fixing $\Omega_\Lambda^\gm=\Omega_\Lambda^\gw$
    and increasing both parameters by 0.03.  The red dot--dashed curve
    corresponds to the red triangular symbol, which is obtained by fixing
    $\overline{\Omega}_\Lambda$ and increasing $\Omega_\Lambda^\gw$ by
    0.03 while decreasing $\Omega_\Lambda^\gm$ by 0.03.}
\end{figure}
\subsection{Weak Gravitational Lensing}

Cosmic shear, due to weak lensing (WL) by large scale structures, has
been detected by several groups~\cite{CS}.  The data set used in our
analysis is from the 75 deg$^2$ Cerro Tololo Inter--American Observatory
(CTIO) lensing survey~\cite{CTIO}.  It covers scales between 1 arcmin
$<\theta<$ 1 deg.  To utilize the WL measurements on small scales, we
take into account nonlinear effects using (i) in the $\Lambda$CDM
case, the nonlinear power spectrum based on the halo model~\cite{NL1};
or (ii) in the QCDM case, the mapping prescription in~\cite{NL2}.

%%%
\section{Estimating Likelihoods}

We use the MCMC package \textit{CosmoMC}~\cite{CosmoMC} to perform our likelihood
analysis.  \textit{CosmoMC} uses \textit{CAMB}~\cite{LCL} to calculate the temperature,
polarization and matter power spectra.  We modify both the \textit{CAMB} and the
MCMC portions to implement the parameter--splitting technique.  In
addition to the dark energy density and EOS parameters
$(\Omega_\Lambda^\gm, \Omega_\Lambda^\gw, w^\gm, w^\gw)$, our
cosmological parameter space includes the baryon density, the Hubble
constant, the reionization optical depth, the scalar spectral index and
amplitude of the primordial power spectrum:
$(\Omega_b h^2, h, \tau, n_s, A_s)$.  When $w^\gw\neq -1$, sound speed
of the dark energy is set as 1 in \textit{CAMB}~\cite{LCL}.  For simplicity, we
assume a flat universe for both geometry and growth parameters.  The
Monte Carlo chains are generated by the Metropolis--Hastings
algorithm~\cite{MH}.  We adopt Gaussian priors of $\Omega_b h^2=0.022\pm
0.002$ from Big Bang nucleosynthesis (BBN)~\cite{BBN} and $H_0=72\pm 8$
km s$^{-1}$Mpc$^{-1}$ from the Hubble Space Telescope (HST) key
project~\cite{HST}.

%%%
\section{Results}

Applying our consistency test to the $\Lambda$CDM model, where the EOS
parameters are fixed as $w^\gm=w^\gw=-1$, the upper panel in
Fig.~\ref{fig:LCDM} shows the marginalized constraints on the
$\Omega_\Lambda^\gw$ \textit{vs.} $\Omega_\Lambda^\gm$ plane.  The
confidence contours follow roughly, but not exactly, the
$\Omega_\Lambda^\gm=\Omega_\Lambda^\gw$ line.  The interesting quantity
in this case is the difference $\Delta\Omega_\Lambda\equiv
\Omega_\Lambda^\gm-\Omega_\Lambda^\gw$, whose normalized probability
distribution is shown in the lower panel of Fig.~\ref{fig:LCDM}.  When
all data are utilized, we find the marginalized constraint
$\Delta\Omega_\Lambda=-0.0044^{+0.0058+0.0108}_{-0.0057-0.0119}$ ($68\%$
and $95\%$ C.L. respectively).  Figure~\ref{fig:LCDM} also shows that CMB
anisotropies, when combined either with galaxy clustering or SNe,
deliver most of the overall constraining power, \textit{i.e.}, having
the narrowest likelihood distributions.

We also find the marginalized constraint on the average
$\overline{\Omega}_\Lambda\equiv(\Omega_\Lambda^\gm+\Omega_\Lambda^\gw)/2$
using all data sets:
$\overline{\Omega}_\Lambda=0.744^{+0.016+0.030}_{-0.015-0.031}$.  The
constraint on the difference is almost three times better than the
constraint on the average.  The CMB contour in Fig.~\ref{fig:LCDM}, even
without the addition of other data, already exhibits this trend.  Let us
therefore focus on understanding this phenomenon in the context of CMB.

As illustrated in Fig.~\ref{fig:dif_avg}, increasing both
$\Omega_\Lambda^\gm$ and $\Omega_\Lambda^\gw$ by the same amount 
(with all the other
cosmological parameters fixed) produces only a small shift of
the predicted $C_\ell$ (blue dashed curve).  However, moving in the
orthogonal direction, \textit{i.e.}, increasing $\Omega_\Lambda^\gw$
while decreasing $\Omega_\Lambda^\gm$, creates a much larger shift (red
dot--dashed curve).  
It appears partial cancellations occur between the shift in
the distance to last scattering (a geometrical quantity) 
and the shift in the sound horizon (which controls fluctuation growth) when
one changes both $\Omega_\Lambda^\gm$ and $\Omega_\Lambda^\gw$
by the same small amount, creating a roughly degenerate direction
along $\Omega_\Lambda^\gm = \Omega_\Lambda^\gw$.
Conversely, the effects of the two different shifts roughly add when one
changes $\Omega_\Lambda^\gm$ and $\Omega_\Lambda^\gw$ in opposite
directions, making $\Delta \Omega_\Lambda$ highly constrained.

\begin{figure}%[htp]
  \centering
  \subfigure{\includegraphics[width=85mm]{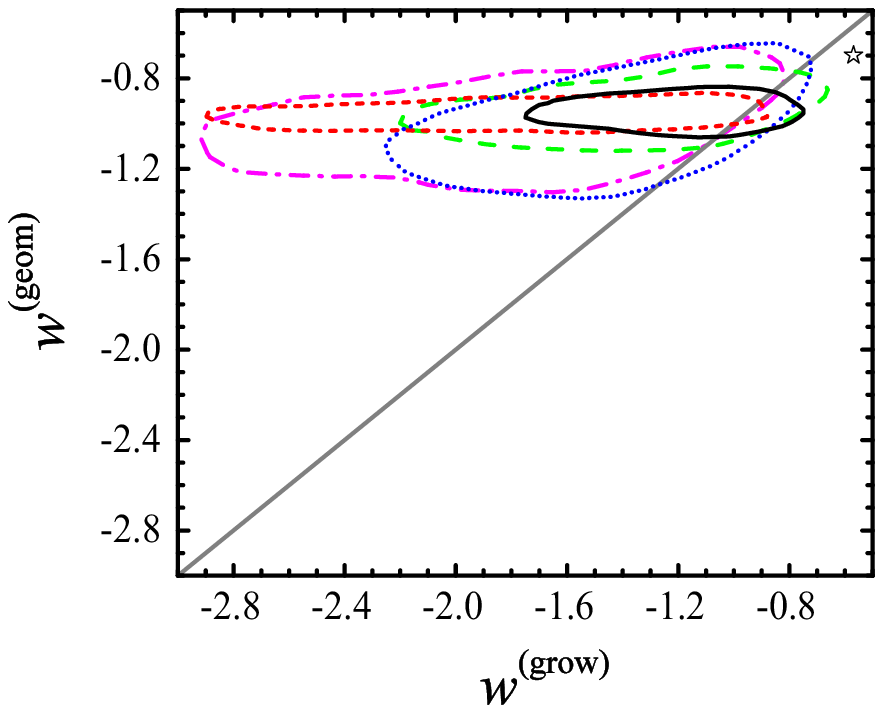}} \\
  \vspace{-1\baselineskip}
  \subfigure{\includegraphics[width=85mm]{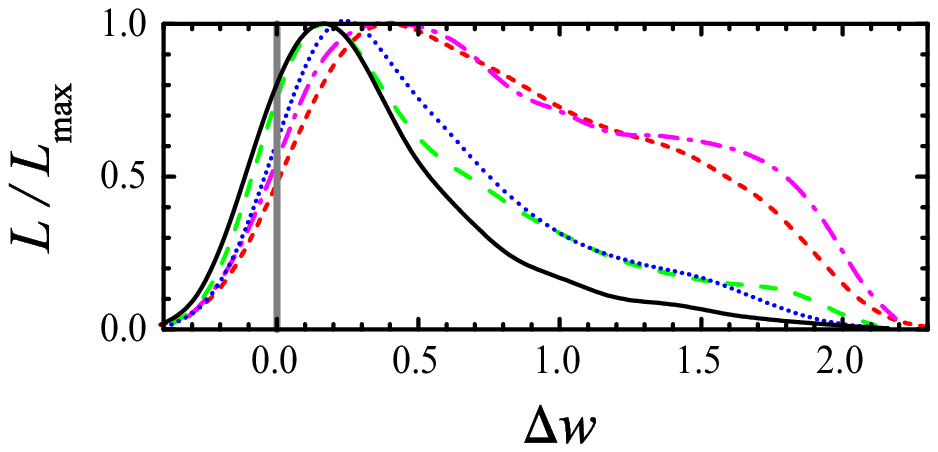}}
  \vspace{-1\baselineskip}
  \caption{\label{fig:QCDM}Joint constraints on $w^\gm$ and $w^\gw$ in
    a QCDM model (\textit{upper panel}) and the normalized likelihood
    distribution of $\Delta w\equiv w^\gm-w^\gw$ (\textit{lower panel}).
    Here the energy density parameters are fixed as $\Omega_{\rm DE}^\gm
    =\Omega_{\rm DE}^\gw$.  The contours and curves show the $68\%$
    confidence limits from the marginalized distributions.  The thick
    gray lines show $w^\gm=w^\gw$.  The data sets used are described in
    the text.  Different contours and curves represent constraints from
    different combinations of the data sets (see legend in
    Fig.~\ref{fig:LCDM}).  The smallest contour and the most narrow
    curve (black solid line) represent constraints from all the data.
    No significant difference is found and deviations are
    constrained to $\Delta w=0.37^{+0.37+1.09}_{-0.36-0.53}$
    ($68\%$ and $95\%$ C.L.).  The star--shaped symbol
    corresponds to the effective $w^\gm$ and $w^\gw$, which approximately
    match the expansion history and the growth history, respectively, of
    a flat DGP model with our best--fit $\Omega_m$.}
\end{figure}

One could argue that in theories of modified gravity constructed to 
explain the late time cosmic acceleration, the growth of fluctuations should
only deviate from GR at late times.  A better approximation of
such theories is perhaps to split the EOS parameter $w$.
We therefore next apply our consistency test to the more general QCDM model.  
The EOS parameters, $w^\gw$ and $w^\gm$, are assumed constant, but are
allowed to vary independently.  In this test, we assume $\Omega_{\rm DE}^\gm 
= \Omega_{\rm DE}^\gw$.  The upper panel in Fig.~\ref{fig:QCDM} shows
the marginalized constraints in the $w^\gw$ \textit{vs.} $w^\gm$ plane~\cite{footnote5}.
We again find that the difference $\Delta w\equiv w^\gm-w^\gw$ is
consistent with zero; deviations are constrained by combining all data
to $\Delta w=0.37^{+0.37+1.09}_{-0.36-0.53}$ (lower panel in
Fig.~\ref{fig:QCDM}).  The average is constrained to be
$\overline{w} \equiv (w^\gm + w^\gw)/2 = -1.13^{+0.18+0.28}_{-0.20-0.55}$.

Figure~\ref{fig:QCDM} shows a long tail towards large
negative values of $w^\gw$, which can be understood as follows.  Density
perturbations can grow significantly only during the matter--dominated
epoch, and as $w^\gw$ becomes more negative, this epoch is longer
(\textit{i.e.}, dark energy domination occurs more recently).  The
extension of the likelihood contours in the large negative direction of
$w^\gw$ reflects the fact that a very recent dark energy domination is
actually acceptable as far as the growth of structure is concerned.
This does not imply the data is consistent with the absence of dark
energy, however.  On the contrary, the data prefer a low $\Omega_m$
which for a flat universe implies the presence of $\Omega_{\rm DE}$.
It is interesting to note that qualitatively, the DGP theory prefers
$w^\gw > w^\gm$~\cite{LSS,footnote1}, a region that is quite restricted
by data already.  In fact, we find that a DGP model with our best--fit
$\Omega_m$, represented effectively by the star--shaped symbol in
Fig.~\ref{fig:QCDM}, lies beyond the $3\sigma$ contour for constraints
from all the data sets; varying $\Omega_m$ in the DGP model within its
$3\sigma$ limits has little effect on the position of the point.  We
also find the upper limits of $w^\gw < -0.97$ at $1\sigma$ and
$w^\gw < -0.80$ at $2\sigma$~\cite{footnote6}.

%%%
\section{Discussions}

Our study reveals no evidence of a discrepancy between the two split
meta--parameters.  The difference is consistent with zero at the
1$\sigma$ level for the $\Lambda$CDM model and 2$\sigma$ level for the
QCDM model.  We find tight constraints from the existing data sets,
especially on the difference between $\Omega_\Lambda$ derived from
growth and $\Omega_\Lambda$ derived from geometry (better than $1\%$).
In other words, the cosmological constant model fits current data very
well.  Current data
do not appear to demand modified gravity theories.  Parameter splitting
can be thought of as a crude way to parametrize the space of these
theories.  As such, our constraints can be viewed as putting
restrictions on modified gravity theories, but the precise constraints
on any particular theory must be worked out on a case by case basis.
The kind of constraints we obtain here are likely to significantly
improve in the future, as the cosmological data improve in quality and
quantity.  The power of future surveys is demonstrated by a calculation
that a Large Synoptic Survey Telescope (LSST)--like survey could
constrain $\Delta w$ to $0.04$, using shear tomography alone, an order
of magnitude better than current constraint from all data
sets~\cite{ZHS,JZ}.

\begin{acknowledgments}
The authors would like to thank Mike Jarvis for providing the CTIO
lensing survey data.  The authors also thank Henk Hoekstra for providing
the Canada--France--Hawaii Telescope Legacy Survey (CFHTLS) data which we
plan to use for future analysis when it is in its final form.  We thank
the WMAP team for making data and the likelihood code public via the
Legacy Archive for Microwave Background Data Analysis (LAMBDA), and
Anthony Lewis and Sarah Bridle for making their MCMC
software~\cite{CosmoMC} available.  The MCMC analyses are performed on
the Columbia Astronomy department computer cluster and the Brookhaven
LSST computer cluster.  This work was supported in part by the DOE under
Contracts No. DE--AC02--98CH10886 and No. DE--FG02--92--ER40699, by the
NSF through Grant No. AST0507161, and by the Initiatives in Science and
Engineering (ISE) Program at Columbia University.
\end{acknowledgments}


\begin{thebibliography}{hello}
\bibitem{footnote1}
Strictly speaking, the DGP theory is not described by a constant
equation of state $w\gm$ for the geometrical distances, nor a constant
$w\gw$ for the fluctuation growth.  However, we find that a model with
suitably chosen values of $w\gm$ and $w\gw$ can approximate the DGP
predictions for the angular diameter distance and the linear growth
rate~\cite{LSS} to within $1.5\%$ for the entire redshift range relevant
for current data (keeping $\Omega_m$ fixed).
\bibitem{footnote2}
As pointed out earlier, there is some freedom in how one carries out the
parameter splitting.  For example, consider the integrated Sachs--Wolfe
term~\cite{dodelson}: $\tilde{S}^{~\rm ISW}_T=e^{-\tau(z)}
[\partial\Psi(k,z)/\partial z-\partial\Phi(k,z)/\partial z]$, where
$\tau$ is the optical depth, $\Psi$ and $\Phi$ are the two scalar metric
perturbations.  One could argue that the integrated optical depth $\tau$
is a ``geometrical'' quantity.  Our point of view is that the optical
depth enters into the problem only through the fluctuation equations.
And in our parameter--splitting scheme, dark energy parameters that
enter into all quantities in the fluctuation equations are in the
``growth'' category.
\bibitem{footnote3}
One subtlety is worth noting.  Existing Boltzmann
codes often output $S_T(k,\eta)\equiv|dz/d\eta|~\tilde{S}_T(k,z)$ as a
function of conformal time $\eta$, rather than redshift.  In carrying
out the geometry--growth split of Eq.~(\ref{eq:CMB}), one must be
careful in noting that the conformal time corresponding to a specific
redshift is not the same for the growth terms (such as $w^\gw$) as for
the geometry terms (such as $w^\gm$).  In other words, changing the
variable of integration from redshift to conformal time,
Eq.~(\ref{eq:CMB}) looks like
$\int_0^{\eta_0}d\eta~S_T(k,\eta)j_\ell[k(\tilde{\eta}_0-\tilde{\eta})]$,
where $\eta$ and $\tilde{\eta}$, while normally equal, should be allowed
to be different under parameter splitting: the requirement is that they
should be chosen to correspond to the same redshift.
\bibitem{footnote5}
The contour of CMB (magenta dot--dashed curve) and the one of CMB
combined with SNe (red short--dashed curve) show a lower bound of
$w^\gw > -2.9$.  This is, however, an artifact as a hard lower limit of
$w^\gw > -3$ is imposed in our MCMC analyses.
\bibitem{footnote6}
Our definition is that the region between the $1\sigma$ ($2\sigma$)
upper limit and the median of the likelihood distribution of the
parameter encloses $68\%/2=34\%$ ($95\%/2=47.5\%$) of the total
likelihood.
\end{thebibliography}
\end{document}